\documentclass{article}
\usepackage[utf8]{inputenc}
\usepackage{authblk}
\usepackage{setspace}
\usepackage[margin=1.25in]{geometry}
\usepackage{graphicx}
\graphicspath{ {./figures/} }
\usepackage{subcaption}
\usepackage{amsmath}
\usepackage{lineno}
\nolinenumbers
\usepackage{booktabs} 
\usepackage{hyperref}
\usepackage{cleveref}
\usepackage{amssymb} 

\usepackage[style=nejm, 
citestyle=numeric-comp,
sorting=none,
backend=bibtex]{biblatex}
\title{The economics of global personality diversity}

\author[a,b,h,*]{Paul X. McCarthy}
\author[h]{Xian Gong}
\author[c]{Marieth Coetzer}
\author[a]{Marian-Andrei Rizoiu}
\author[d]{Margaret L. Kern}
\author[f]{John A. Johnson}
\author[j]{Richard Holden}
\author[c,e]{Fabian Braesemann}

\affil[a]{\footnotesize The Data Science Institute, University of Technology Sydney, Australia}
\affil[b]{\footnotesize School of Computer Science and Engineering, UNSW Sydney, Australia}
\affil[c]{\footnotesize Oxford Internet Institute, University of Oxford, UK}
\affil[d]{\footnotesize Centre for Wellbeing Science, University of Melbourne, Australia}
\affil[e]{\footnotesize Einstein Center Digital Future, Berlin, Germany}
\affil[f]{\footnotesize Department of Psychology, Pennsylvania State University, USA}
\affil[j]{\footnotesize School of Economics, UNSW Sydney, Australia}
\affil[h]{\footnotesize League of Scholars, Sydney, Australia}
\affil[*]{\footnotesize Correspondence to: paul@leagueofscholars.com}

\date{}

\begin{document}

\maketitle

\begin{abstract}
Understanding why national economic performance differs remains a central question in economics. Although institutions and demographic diversity are established correlates of economic development, the contribution of within-population personality diversity has received limited attention. We introduce the Global Personality Diversity Index ($\Psi$-GPDI), a measure of psychological heterogeneity derived from assessments of 760,242 individuals across 135 countries using 30 facets of the Big Five personality traits. We show that most variation in these facets occurs within rather than between countries. Across countries with available economic data, personality diversity is positively associated with GDP per person employed, accounting for 19.9\% of the variance in an adjusted single-predictor model. Including $\Psi$-GPDI alongside institutional quality and immigrant diversity increases the model's explanatory power by approximately 5.7\%. Personality diversity is also weakly correlated with immigrant diversity, suggesting that the two measures capture distinct characteristics of national populations. These findings identify psychological heterogeneity as a potentially important and previously underexamined dimension of national human capital. Although the observational design does not establish causality, the results motivate further investigation of how personality diversity interacts with institutions, occupational specialization, and economic development.
\end{abstract}


Team diversity enhances performance by fostering innovation and improving problem-solving across various fields. A comprehensive analysis of over 9 million collaboratively authored research papers involving about 6 million scientists found that diversity in ethnicity, age, gender, and institutional affiliation all positively correlate with research impact, with ethnic diversity associated with up to 40\% more citations~\cite{alshebli2018preeminence}.

While the term “diversity” today often emphasizes categorical group differences such as ethnicity, gender, and culture, this approach can sometimes overlook the critical variations within these categories, potentially implying homogeneity that does not exist. Individual differences, encompassing both personality and intellectual abilities, represent a fundamental level of diversity. By focusing on personality diversity, this study explores a dimension of human variability that transcends group categorizations.

The significance of personality diversity is evident not only in research teams but also in high-growth firms. A large-scale study involving over 21,000 startups found that those with diverse founder personalities were 8–12 times more likely to succeed, underscoring the role of psychological differences in innovation and economic performance~\cite{mccarthy2023impact}.

Similarly, diversity within corporate boards has been shown to significantly enhance innovation at the organizational level. An analysis of 4,448 firms and 43,639 directors found that companies with more diverse boards not only generated a higher volume of patents but also developed innovations of greater impact, as indicated by increased citation frequencies~\cite{chen2023board}. 

The advantages of diversity are well established at the micro level, in research teams, startups, and corporations. Far less is known about its effect on economic performance at the macro level, even though income disparities between countries remain a central question in economic research. Acemoglu, Johnson, and Robinson~\cite{acemoglu2001colonial} highlighted how inclusive institutions promote prosperity, whereas extractive institutions hinder economic growth. Nonetheless, institutional factors alone do not fully explain these income disparities, underscoring the need to examine additional dimensions that influence economic outcomes. 

Beyond institutional frameworks, immigrant diversity emerges as a second significant factor influencing economic performance through various forms of diversity, including birthplace diversity~\cite{alesina2016birthplace}, ethnic diversity~\cite{Alesina2005}, and genetic variation~\cite{Ashraf2013}.

A third, underexplored dimension is personality. Recent research by Shleifer and colleagues~\cite{categories2008,bordalo2012salience,bordalo2013asalience,bordalo2013bsalience,bordalo2016competition,bordalo2020memory,bordalo2023memory,bordalo2024cognitive} aims to delve into individual cognitive processes and their implications for decision-making. Similarly, Akerlof, Holden, and Li~\cite{akerlof2024} have developed models of human cognition and reasoning. And on the economic geography front, Garretsen et al. have shown links between regional economic growth and the presence of specific personality features such as Conscientiousness~\cite{garretsen2019relevance}. However, the relationship between individual-level personality traits and aggregate macroeconomic outcomes has received limited attention in economic studies to date.

Building on the exploration of individual personality traits, a largely unexamined dimension is national personality diversity.  While studies like McCrae and Terracciano~\cite{mccrae2005personality} have focused on differences in average personality traits between countries to understand cross-cultural variations, they did not examine the economic implications of these traits nor the variability of personality within nations. This gap makes the case for exploring how within-nation personality diversity, rather than just average national traits, might influence economic performance.

We hypothesize that personality diversity could support a nation's innovation and productivity. Countries with diverse personality traits may better handle challenges and foster creativity. Our study investigates whether national personality diversity correlates with higher GDP per person employed. We posit that psychologically diverse populations contribute a wider array of perspectives, skills and problem-solving approaches, which may support economic performance.

To test this hypothesis, we conducted a multifactor analysis incorporating known variables influencing national income: immigrant diversity rates and expropriation risk (institutional strength proxy), as highlighted by Acemoglu et al.~\cite{acemoglu2001colonial}. Including these factors, we test whether personality diversity explains additional variance in economic performance. Inspired by the Fama-French three-factor model, we seek to identify factors that complement known measures, particularly institutional strength and immigrant diversity. 

We introduce the Global Personality Diversity Index ($\Psi$-GPDI), a metric quantifying national personality diversity, constructed from data on 760,242 individuals across 135 countries using 30 facets of the Big Five personality traits~\cite{costa1992four}. Calculating trait variability within each country provides a standardized measure, allowing examination of correlations between $\Psi$-GPDI and economic indicators.

If personality diversity is associated with economic performance, an important next question is under which conditions individuals with different dispositions can contribute their distinct capacities, for example through education, workforce participation, and occupational specialization. This perspective complements institutional and cultural explanations of national prosperity by shifting attention from average personality profiles to the diversity of personalities within populations.

\section*{Results}

To explore global personality diversity, we analyzed a comprehensive dataset comprising responses from 760,242 individuals across 135 countries collected between 2001 and 2022. Each participant completed the 300-item IPIP-NEO personality inventory, providing detailed assessments of the Big Five personality traits and their 30 facets. The dataset includes basic demographic information such as country of residence, age, and gender, allowing us to examine personality patterns across different nations and demographic groups.

\subsection*{Within-country personality diversity}

\begin{figure}[t!]
\centering
  {\includegraphics[width=0.7\linewidth, keepaspectratio]{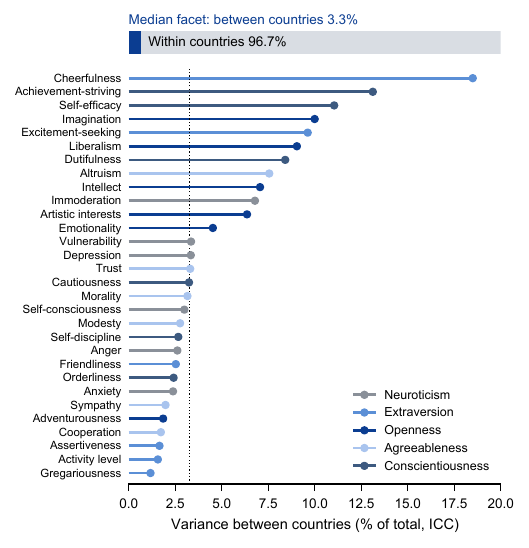}}
  \caption{\textbf{$|$ Most variation in personality facets lies within countries rather than between them.} For each of the 30 Big Five facets, the horizontal line and dot show the share of total variance attributable to differences between countries (intraclass correlation, ICC, expressed as a percentage). Facets are ordered by this share and colored by Big Five domain. The dotted vertical line marks the median facet (3.3\%); the bar at the top shows the corresponding split for the median facet, with 3.3\% of the variance between countries and 96.7\% within countries. Cheerfulness has the largest between-country share and Gregariousness the smallest. Because national averages capture only a small part of personality variation, comparing countries by their average profiles (SI Appendix, Fig.~\ref{fig:dendrogram}) misses most of it. This motivates measuring the diversity of personalities within each country (Fig.~\ref{fig:pairwise_distributions}).}\label{fig:icc}
\end{figure}

\begin{figure}[t!]
\centering
  {\includegraphics[width=0.6\linewidth, keepaspectratio]{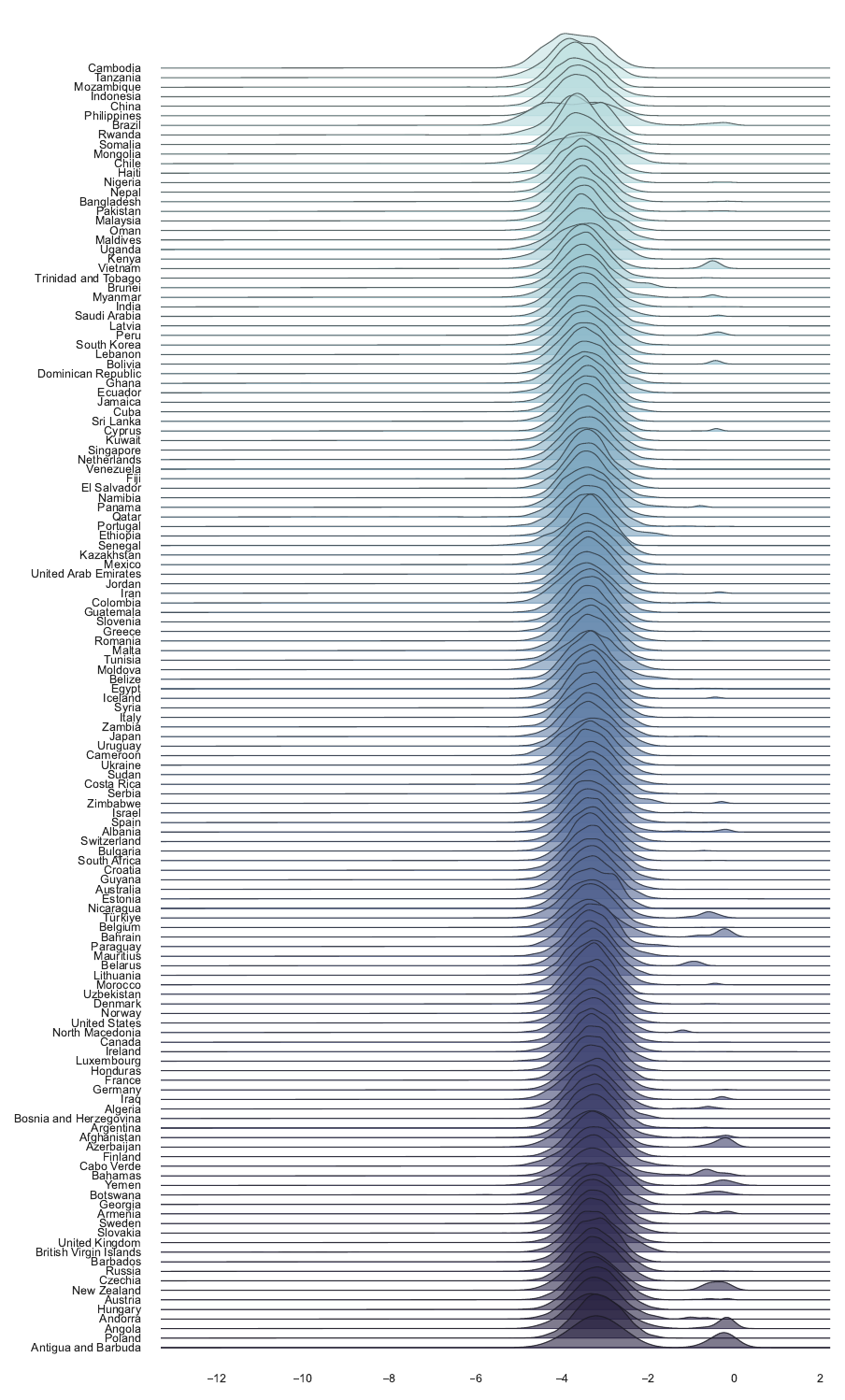}}
  \caption{\textbf{$|$ In some countries, people are more alike in personality than in others.} Ridgeline plot showing, for each country, the distribution of log-transformed pairwise cosine distances between individuals' 30-dimensional personality profiles. Each distribution shows how different two randomly chosen residents of a country are: further left means more alike, further right less alike. Countries are ordered by $\Psi$-GPDI, from lowest (top) to highest (bottom). The Global Personality Diversity Index ($\Psi$-GPDI) is derived from these distributions, as the inverse of the absolute median of each (see \hyperref[sec:methods]{Materials and Methods}), so a rightward shift corresponds to a higher index. Countries differ not only in this central tendency but also in the dispersion and shape of their distributions, with several showing secondary peaks or shoulders. A single index value therefore captures the typical level of personality diversity in a country but not all of its internal structure. SI Appendix, Fig.~\ref{fig:ks_heatmap} tests whether these national distributions differ significantly from one another (93.53\% of the 9,045 country pairs differ at $p < 0.05$).}\label{fig:pairwise_distributions}
\end{figure}

Comparing countries by their median personality profiles reveals some regional structure, but hierarchical clustering also links geographically distant countries, and none of its three top-level clusters corresponds to a single world region (SI Appendix, Fig.~\ref{fig:dendrogram}). Here we focus instead on personality heterogeneity within each country. A decomposition of the variance of each of the 30 facets into between-country and within-country components shows why this second perspective matters (Fig.~\ref{fig:icc}). For the median facet, differences between countries account for only 3.3\% of the total variance (intraclass correlation, ICC), leaving 96.7\% within countries. The between-country share varies across facets, from roughly 1\% for Gregariousness to about 18\% for Cheerfulness, and it reaches about 10\% or more for only four facets (Cheerfulness, Achievement-striving, Self-efficacy and Imagination). Even for these facets, most of the variation in personality lies between individuals within the same country rather than between national averages. We quantify this heterogeneity using pairwise distances between individuals' 30-dimensional personality profiles, involving over 3.5 billion pairs after subsampling (see \hyperref[sec:methods]{Materials and Methods}). These distances form the basis of the Global Personality Diversity Index ($\Psi$-GPDI), providing a distributional
representation of within-country personality diversity (Fig.~\ref{fig:pairwise_distributions}). The resulting distributions vary substantially across the 135 countries, not only in central tendency but also in dispersion and shape, and several (such as Brazil, New Zealand, and Angola) show secondary peaks or shoulders. Within-country personality structure therefore cannot always be summarized by a single central value.

Two countries may have similar median personality profiles while differing substantially in the heterogeneity of individuals around those profiles. The between- and within-country analyses therefore distinguish the \textit{composition} of national personality profiles from the \textit{diversity} of personalities within national populations.

\subsection*{In-country personality archetypes}

Within-country diversity is also structured. As an illustration, hierarchical clustering of a random 10\% sample of U.S. respondents (n = 44,678) identifies eight archetypal combinations of facets, which we call the PLATYPUS typology (SI Appendix, Fig.~\ref{fig:us_heatmap} and Table~\ref{tab:platypus}). Two archetypes, Pathfinders and Leaders, contain roughly twice the share of individuals predicted to have an entrepreneurial personality profile as the other six (about 17\% versus 9\%; SI Appendix, Fig.~\ref{fig:prob}), indicating that personality types within a population differ in economically relevant capacities. Because the clustering is applied to a two-dimensional t-SNE embedding, we treat these archetypes as descriptive (SI Appendix).

\subsection*{The Global Personality Diversity Index ($\Psi$-GPDI)}
To capture and quantify in-country personality diversity, we developed the Global Personality Diversity Index ($\Psi$-GPDI), a measure based on the pairwise cosine distance between 30-dimensional personality trait vectors derived from the IPIP-NEO inventory. The $\Psi$-GPDI incorporates tens of millions of pairwise comparisons within each country, and offers a robust assessment of personality diversity at the national level (see \hyperref[sec:methods]{Materials and Methods}).

\begin{figure*}[t!]
\centering
  {\includegraphics[width=0.9\linewidth, keepaspectratio]{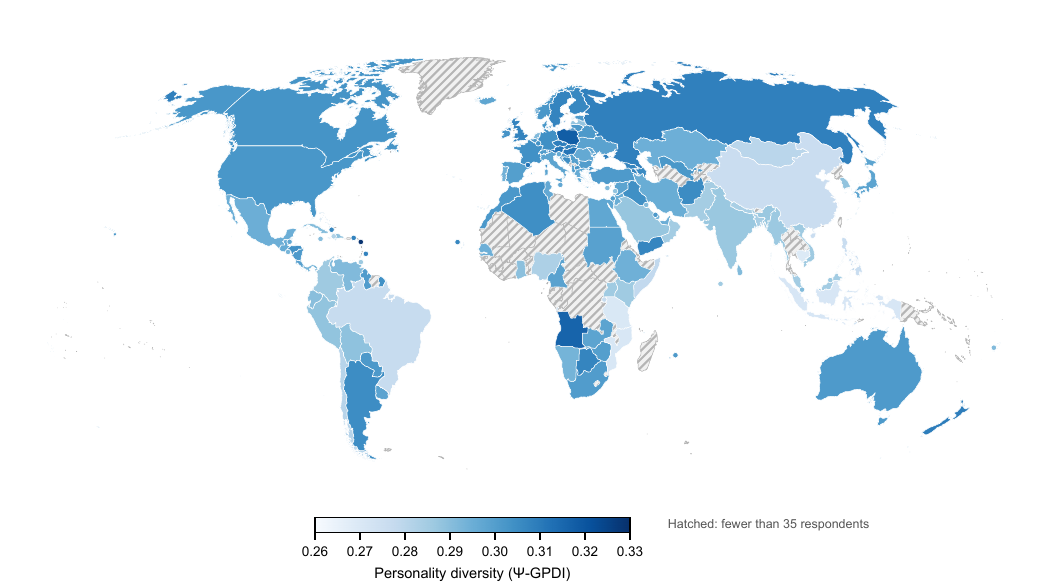}}
  \caption{\textbf{$|$ Personality diversity varies across countries and does not simply follow world regions.} World map of the Global Personality Diversity Index ($\Psi$-GPDI) for the countries in the sample, with darker shades indicating greater within-country personality diversity (color scale from 0.26 to 0.33). Values tend to be higher in Europe, North America and Australasia and lower in China, much of Southeast Asia and parts of eastern Africa, yet both higher and lower values occur within most regions. Small countries and island states are drawn as dots so that they remain visible. Hatched areas denote countries with fewer than 35 respondents, for which the index is not computed (see \hyperref[sec:methods]{Materials and Methods}).}\label{fig:map}
\end{figure*}

Fig.~\ref{fig:map} maps the resulting country-level $\Psi$-GPDI values. The index spans a narrow numerical range, from about 0.26 to 0.33, yet its geographic distribution is far from uniform. Values tend to be higher across much of Europe, Canada and the United States, Russia, Australia and New Zealand, as well as in Argentina and parts of southern Africa, with Poland and Angola among the highest of the larger countries. Values tend to be lower in China and much of Southeast Asia, in Brazil and in parts of eastern Africa. Both higher and lower values nevertheless occur within most world regions, so that personality diversity does not simply follow regional groupings. Countries with fewer than 35 respondents, most of them in Africa, are excluded from the index and shown hatched.

Distributional shape also varies (Fig.~\ref{fig:pairwise_distributions}). Some countries, such as New Zealand and South Africa, show multimodal distributions, which may reflect distinct subgroups within their populations, whereas countries such as Japan and Finland show single, sharp peaks. Pairwise Kolmogorov-Smirnov tests (SI Appendix, Fig.~\ref{fig:ks_heatmap}) indicate that 93.53\% of the 9,045 country pairs significantly differ. Because pairwise distances within a country share individuals, they are not independent, and with samples of this size even small differences reach significance. We therefore treat these tests as descriptive evidence that national distributions are distinct rather than as formal inference.

\begin{figure}[t!]
\centering
  {\includegraphics[width=0.8\linewidth, keepaspectratio]{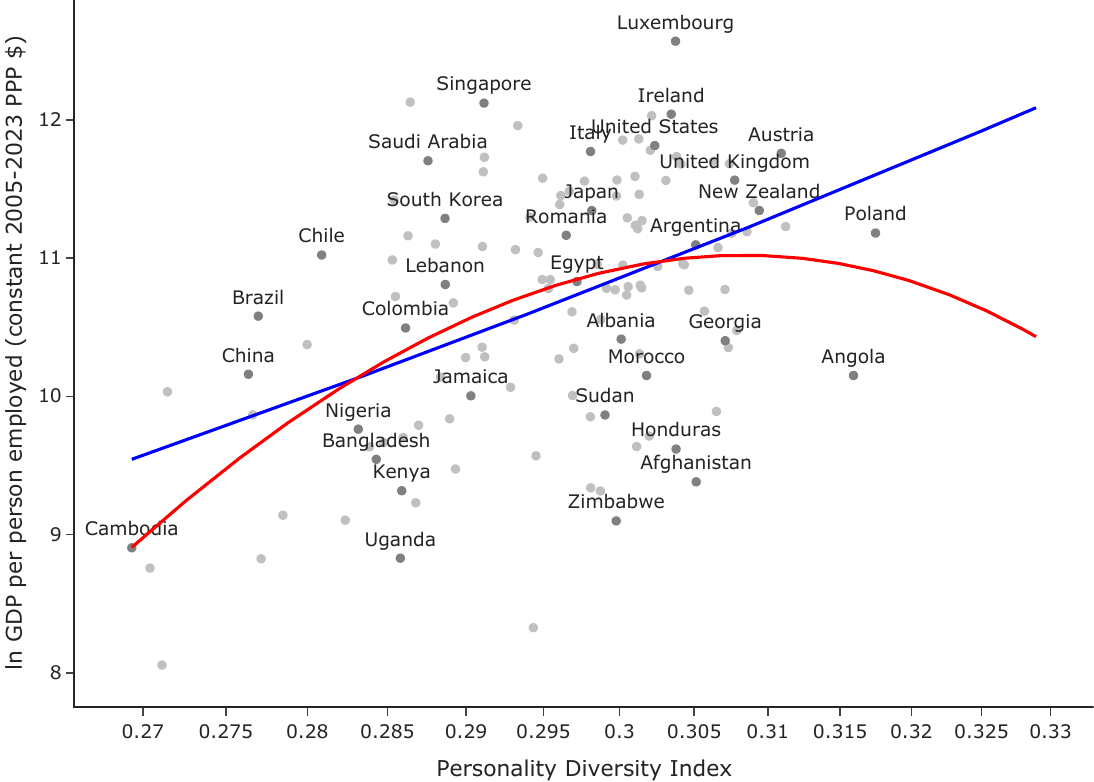}}
  \caption{\textbf{$|$ Countries with greater personality diversity have higher output per worker, with a plateau at the highest levels of diversity.} Scatter plot showing the relationship between $\Psi$-GPDI and GDP per person employed for 129 countries, averaged over 2005--2023 and expressed in constant 2021 PPP international dollars. Values on the vertical axis are natural logarithms. The blue line is a linear fit (unadjusted $R^2 = 20.87\%$) and the red curve a quadratic fit (unadjusted $R^2 = 24.41\%$). The association is positive across most of the observed range (about 0.27 to 0.31) and flattens above it, where countries such as Angola combine high diversity with modest income. The relationship is cross-sectional and does not establish causation. Table~\ref{tab:granger_tab} and SI Appendix, Fig.~\ref{fig:corr} test whether it persists alongside institutional quality and immigrant diversity. For legibility, only a subset of countries across the income range is labeled; all 129 are plotted.}\label{fig:regression}
\end{figure}

\subsection*{$\Psi$-GPDI correlates with national income}
The analysis reveals a significant relationship between $\Psi$-GPDI and national income across 129 countries, measured as GDP per person employed (Fig.~\ref{fig:regression}). Six countries (Venezuela, Cuba, Yemen, Andorra, Antigua and Barbuda, and the British Virgin Islands) lack World Bank estimates of GDP per person employed and are excluded from this section. At moderate levels of $\Psi$-GPDI (0.27–0.31), a strong positive correlation is observed, with high-income nations such as Luxembourg, Ireland, and the United States achieving particularly strong economic performance. Personality diversity is thus measurably associated with national output per worker across a broad global sample. Linear and quadratic models explain 20.87\% and 24.41\% of the variance in GDP per person employed, respectively (unadjusted R²). Above $\Psi$-GPDI values of about 0.31 the relationship plateaus or slightly declines, as in Angola, where high personality diversity coincides with modest output per worker; this may indicate diminishing returns to diversity or the influence of other structural factors. The multifactor analysis below reports adjusted $R^2$, which corrects for the number of predictors, for countries with complete data on all three predictors.

\subsection*{Multifactor model results}
A multifactor analysis was conducted to assess the contribution of $\Psi$-GPDI to national income alongside expropriation risk (a proxy for institutional quality; higher values indicate more secure property rights) and immigrant diversity (see Table~\ref{tab:granger_tab}). While the preceding correlation analysis included 129 countries, the multifactor analysis was restricted to 116 countries due to missing expropriation risk data for 13 countries (Afghanistan, Nepal, Bosnia and Herzegovina, Georgia, North Macedonia, Cambodia, Mauritius, Barbados, Cabo Verde, Belize, Maldives, Fiji, and Rwanda). Individually, expropriation risk exhibited the strongest association with GDP (adjusted $R^2 = 50.8\%$), followed by immigrant diversity ($23.1\%$) and $\Psi$-GPDI ($19.9\%$). Consistent with these results (see SI Appendix, Fig.~\ref{fig:corr}), GDP was significantly correlated with expropriation risk ($r = 0.716$), immigrant diversity ($r = 0.488$), and $\Psi$-GPDI ($r = 0.453$).

Combining these factors substantially increased explanatory power. Expropriation risk and immigrant diversity jointly explained 57.1\% of the variance in GDP, while expropriation risk and $\Psi$-GPDI explained 56.1\%. The combination of immigrant diversity and $\Psi$-GPDI accounted for 40.0\%, and the full model incorporating all three factors accounted for 62.8\%. An increase of 5.7\% over the model with expropriation risk and immigrant diversity alone. SI Appendix, Fig.~\ref{fig:gpdi_residual} shows the partial relationship between $\Psi$-GPDI and GDP conditional on the other two factors.

\begin{table}[!ht]
\centering
\caption{\textbf{$|$ Contributions of Expropriation, Immigrant Diversity, and Personality to GDP.}  This table shows the adjusted R-squared values for individual and combined factors across 116 countries (see \hyperref[sec:methods]{Materials and Methods}). Asterisks denote statistical significance: *** $p<0.001$.}\label{tab:granger_tab}
  \begin{tabular}{lc}
    \toprule
    Models & Adjusted R-squared \\
    \midrule
    GDP \textasciitilde{} Expropriation risk & 50.8\% (***) \\
    GDP \textasciitilde{} immigrant diversity & 23.1\% (***) \\
    GDP \textasciitilde{} $\Psi$-GPDI & 19.9\% (***) \\
    GDP \textasciitilde{} Expropriation risk + immigrant diversity & 57.1\% (***) \\
    GDP \textasciitilde{} Expropriation risk + $\Psi$-GPDI & 56.1\% (***) \\
    GDP \textasciitilde{} immigrant diversity + $\Psi$-GPDI & 40.0\% (***) \\
    GDP \textasciitilde{} Expropriation risk + immigrant diversity + $\Psi$-GPDI & 62.8\% (***) \\
    \bottomrule
  \end{tabular}
\end{table}

\section*{Discussion}
Geographic variation in personality has been documented both within and across countries. Cross-national studies covering up to 56 countries have identified substantial variation in Big Five personality traits~\cite{schmitt2007geographic,mccrae2005personality,allik2004toward}. Within countries, personality also exhibits pronounced spatial structure. Studies have identified distinct regional personality profiles in the United States~\cite{huggins2021behavioral}, associations between personality and physical geography and ecology~\cite{gotz2020physical}, and geographic clustering of traits such as neuroticism and conscientiousness in the United Kingdom~\cite{garretsen2019relevance}.

Previous cross-national clustering based on broad Big Five domains identified meaningful similarities among 36 countries~\cite{allik2004toward}. We extend this approach by representing each country using 30 Big Five personality facets and analyzing a substantially larger sample of 135 countries (SI Appendix, Fig.~\ref{fig:dendrogram}). Our findings distinguish three related but conceptually different dimensions of personality variation: differences in aggregate personality profiles between countries, heterogeneity in personality archetypes within countries, and within-country personality diversity quantified by $\Psi$-GPDI.

Geographic variation in personality may also be economically relevant. Regional differences in traits such as conscientiousness and openness have been associated with entrepreneurial activity and economic development~\cite{obschonka2013regional,garretsen2019relevance}, while broader regional psychological profiles have been linked to differences in economic performance~\cite{garretsen2019relevance,huggins2021behavioral}. Related evidence at the global level indicates that economic preferences, particularly patience, are strongly associated with national economic outcomes~\cite{falk2018global}.

\subsection*{Within-country variation}

Personality variation within countries may be economically relevant because different social and occupational roles are associated with distinct combinations of psychological characteristics. Previous research on 21,187 startups found that, although particular personality traits were associated with entrepreneurship, founding teams with greater personality diversity were substantially more likely to succeed~\cite{mccarthy2023impact}. This finding suggests that economic performance may depend not only on particular traits but also on the complementarity of different personality profiles.

This interpretation is consistent with evidence linking personality to occupational choice and performance. A large-scale analysis of 128,279 individuals across 3,513 occupations demonstrated systematic associations between personality profiles and occupational roles~\cite{kern2019social}. These findings extend a long tradition of research on personality-occupation fit, originating with Strong's work on vocational interests~\cite{strong1943vocational}, subsequently formalized through Holland's theory of vocational personality types~\cite{holland1997making}, and further connected to the Big Five framework by Costa and McCrae~\cite{costa1984personality}. Collectively, this literature suggests that occupational systems benefit from heterogeneous psychological profiles suited to different functional roles.

The U.S. archetypes illustrate this heterogeneity (SI Appendix, Fig.~\ref{fig:us_heatmap} and Table~\ref{tab:platypus}): groups are distinguished by combinations of facets rather than single traits, and their predicted entrepreneurial propensity differs. This does not imply that any archetype is more economically valuable than another. Entrepreneurship and innovation coexist with occupations in science, education, health, administration, and regulation that draw on different psychological attributes, so the typology describes how heterogeneous personality profiles may contribute complementary capabilities within a population.

\subsection*{Validation of $\Psi$-GPDI}

Because $\Psi$-GPDI is derived from self-reported IPIP-NEO responses, we compared it with an independent benchmark: the country-level standard deviations of observer-rated personality reported by McCrae and Terracciano~\cite{mccrae2005personality}, based on 12,156 participants across 51 countries. Across the 48 countries present in both datasets, the two measures are positively associated ($R^2 = 13.26\%$ in SI Appendix, Fig.~\ref{fig:index_validation}). Given their differences in instrument, rater, and statistic, we regard this as modest convergent evidence rather than definitive validation: countries that are more diverse by $\Psi$-GPDI tend to show greater personality variation in observer ratings, but the two measures are not interchangeable.

A central finding of this study is the positive cross-national association between personality diversity and output per worker (Fig.~\ref{fig:regression}). The better fit of the quadratic specification suggests that the association may flatten at the highest levels of diversity. Because the data are cross-sectional, they neither establish nor rule out a causal link; testing whether personality diversity contributes to economic performance will require longitudinal or quasi-experimental evidence.

Supplementary analyses indicate that this relationship is not confined to a single regional or economic grouping. Among OECD and G20 countries (SI Appendix, Fig.~\ref{fig:combined_regression_a}), the linear and quadratic specifications explain 32.47\% and 41.53\% of the variance, respectively. A positive association is also observed among the 24 African countries included in the analysis (SI Appendix, Fig.~\ref{fig:combined_regression_b}; linear and quadratic $R^2$ of 44.85\% and 46.05\%). These results suggest that the association between personality diversity and economic outcomes is observable across substantially different socioeconomic contexts, while its magnitude and functional form may vary between groups. 

The multifactor analysis positions personality diversity as a complementary explanatory factor rather than a substitute for established institutional and demographic correlates of national income: adding $\Psi$-GPDI to expropriation risk and immigrant diversity raises the adjusted $R^2$ by 5.7\% (Table~\ref{tab:granger_tab}). The two comparison variables were chosen because they represent the institutional and diversity-based explanations that $\Psi$-GPDI is intended to complement. They are not an exhaustive set of controls: education, health, geography, demographic structure, and regional history could also confound the association, so the incremental association should not be read as causal.

The near-zero association between $\Psi$-GPDI and immigrant diversity ($r = 0.081$) reflects a difference in construct rather than an absence of structure. Birthplace diversity is a categorical composition measure, indexing the probability that two randomly selected foreign-born residents originate in different countries, weighted by the migrant share of the population. It is therefore computed over a subpopulation that in most countries does not exceed 15\% of residents, and it treats every pair of origin countries as equally distant. Our own results indicate that this assumption does not hold for personality: the pairs of countries with the most similar national personality profiles in SI Appendix, Fig.~\ref{fig:dendrogram} include Spain and Portugal, and Austria and Germany, yet migrants from either pair contribute to measured birthplace diversity exactly as much as migrants from countries at opposite ends of the distribution. $\Psi$-GPDI, by contrast, is a continuous distance measure defined across the entire resident population. Two further considerations point the same
way. Within-country variation in Big Five traits is substantially larger than between-country variation in mean trait levels~\cite{terracciano2005national}, a pattern that our own data reproduce at the facet level, where the median facet has only 3.3\% of its variance between countries (Fig.~\ref{fig:icc}), so variation in national origin need not translate into variation in personality. And migrants are not random draws from their origin populations, so migration may introduce residents from many origins whose personality profiles are comparatively similar. Taken together, these considerations indicate that $\Psi$-GPDI captures a dimension of national population structure that is distinct from origin-based diversity, consistent with its low variance inflation factor in the multifactor model.

$\Psi$-GPDI is, by contrast, modestly associated with institutional quality ($r = 0.320$). The direction of this relationship cannot be established from cross-sectional data, and at least three interpretations are consistent with it. Institutional conditions may shape the range of dispositions that individuals are able to express, if secure rights and constrained state power permit a wider variety of occupational, social and political choices. Alternatively, maintaining independent courts, bureaucracies and other institutions that rely on adversarial or specialized roles may itself depend on a population with heterogeneous psychological profiles. A third possibility is that development, education and urbanization contribute to both. The association is moderate rather than strong, and the variance inflation factor of 1.12 indicates that $\Psi$-GPDI remains largely independent of institutional quality within the model. We regard the relationship between psychological diversity and institutional structure as an open question that longitudinal or subnational data would be better placed to address.

Taken together, these findings suggest that population-level personality diversity represents a potentially relevant dimension of national human capital. A heterogeneous distribution of personality profiles may provide a broader range of behavioral capacities for entrepreneurship, collaboration, creativity, problem solving, and specialized occupational roles. However, diversity alone does not guarantee economic benefits. Its potential contribution is likely to depend on institutional conditions that enable individuals with different psychological profiles to participate effectively and interact across social and occupational boundaries. Accordingly, the results motivate further investigation of personality diversity within broader frameworks of human capital, immigrant diversity, innovation, and economic development rather than implying direct policy effects from the observed associations.

\subsection*{Limitations and future directions}
Several limitations should be considered when interpreting these findings. First, population-level personality diversity does not necessarily imply effective interaction among individuals with different personality profiles. Social sorting may lead individuals with similar characteristics to associate disproportionately with one another, potentially limiting the benefits of aggregate diversity. Ma et al.~\cite{ma2024threats}, for example, identify perceived personal control as one factor associated with social sorting. Future research should therefore examine not only the level of personality diversity within populations but also the social and organizational structures through which heterogeneous individuals interact. 

Second, the study is subject to potential sampling and measurement biases. The dataset comprises 760,242 voluntary participants who completed an English-language version of the IPIP-NEO inventory over more than two decades. Consequently, the sample may disproportionately represent individuals with sufficient English proficiency, internet access, and interest in personality assessment. The sample is also younger than the global population, with a median age of 22 compared with a global median of approximately 28.2 years in 2010. These characteristics should be considered when generalizing the results to national populations.

Nevertheless, the sample exhibits meaningful correspondence with external demographic benchmarks. The age distribution is positively correlated with the global age distribution reported by the United Nations for 2010 ($r = 0.64$; SI Appendix, Fig.~\ref{fig:age_population}), while country-level sample sizes are also correlated with national population sizes ($r = 0.54$; SI Appendix, Fig.~\ref{fig:country_population}). These associations indicate broad demographic coverage but do not establish representativeness within countries. Cross-national differences in the age, gender, education, internet access, or English proficiency of respondents could therefore affect $\Psi$-GPDI. Robustness to reweighting by age and gender, to minimum country sample size, and to multilingual or externally representative samples remains to be tested.

Third, $\Psi$-GPDI is one of several possible measures of multivariate dispersion. Cosine distance emphasizes differences in the shape of personality profiles rather than in overall score levels, whereas alternatives such as Euclidean distance on standardized facets or generalized variance weight these components differently. Whether country rankings and the economic associations are robust to such alternatives is an important test.

Finally, the relationships between $\Psi$-GPDI and economic outcomes are observational and cross-sectional. They should therefore be interpreted as associations rather than evidence that personality diversity causes higher national income. Economic development, immigrant diversity, institutions, education, cultural conditions, and personality diversity may influence one another through multiple pathways. Longitudinal analyses, quasi-experimental designs, and individual-level linkage between personality, mobility, occupation, and economic outcomes would provide stronger tests of these mechanisms. Future research could also examine whether changes in personality diversity predict subsequent economic performance and under which institutional conditions diversity is most strongly associated with innovation, productivity, and resilience. In addition, GDP per person employed averaged over 2005--2023 measures the level of labor productivity rather than its growth, so the results do not speak directly to innovation or growth dynamics.

\section*{Materials and Methods}
\subsection*{Dataset description}

The IPIP-NEO personality inventory is a comprehensive personality assessment employing 300 items from the International Personality Item Pool (IPIP), a public domain repository developed by Dr. Lewis R. Goldberg~\cite{goldberg1999broad}. It mirrors the structure of the commercial NEO Personality Inventory (NEO PI) by Costa and McCrae~\cite{costa1992four}. The IPIP-NEO assesses the Big Five personality traits along with 30 specific facets. The well-established reliability and validity of the IPIP-NEO have been described by Johnson~\cite{johnson2014measuring}.

The current study combines data collected by Johnson~\cite{johnson2014measuring} from 307,313 respondents who completed the IPIP-NEO between 2001 and 2011, available in the public domain at~\url{https://osf.io/tbmh5/}, with 466,401 further responses collected between 2011 and 2022, giving 773,714 records in total. We excluded 13,472 records from countries with fewer than 35 respondents, as $\Psi$-GPDI is estimated from within-country pairwise distances and is unstable at very small sample sizes. The analysis dataset comprises 760,242 individuals across 135 countries. The psychometric properties of measures from the International Personality Item Pool have been extensively validated~\cite{goldberg2006international}. Methods for the U.S. personality archetype analysis (t-SNE embedding, Ward clustering and selection of the eight-cluster solution) are described in SI Appendix.

\subsection*{Construction of $\Psi$-GPDI}

The dataset contains 300 items on 30 personality facets, with 10 different items corresponding to one personality facet. All items use a five-point Likert scale. We sum the scores of the 10 items to give a raw score for each facet, ranging from 10 to 50. Our input matrix therefore comprises \(N=760,242\) individuals from 135 countries, each represented by a 30-dimensional vector corresponding to distinct personality facets.

Let \(\mathbf{X} \in \mathbb{R}^{N \times 30} \) denote the matrix of personality vectors, where each row \(\mathbf{x}_i\) corresponds to an individual \(i\). The known country affiliation of each individual allows us to partition \(\mathbf{X}\) into subsets \(\mathbf{X}_c\) for each country \(c\) with \(N_c\) individuals. We use Eq.~\ref{equ1} to compute the pairwise cosine distance between all unique pairs of individuals within the country. The cosine distance between two vectors \(\mathbf{x}_i\) and \(\mathbf{x}_j\) is defined as:
\begin{equation} \label{equ1}
d(\mathbf{x}_i, \mathbf{x}_j) = 1-\frac{\mathbf{x}_i \cdot \mathbf{x}_j}{\parallel \mathbf{x}_i \parallel \cdot \parallel \mathbf{x}_j \parallel}
\end{equation}
where \(\cdot\) denotes the dot product and \(\parallel \cdot \parallel\) denotes the Euclidean norm. A smaller cosine distance indicates greater similarity between two personality profiles, while a larger cosine distance indicates greater dissimilarity. This computation yields a set \(\mathbf{D}_c\) of \(\begin{pmatrix} N_c\\ 2\\ \end{pmatrix}=\frac{N_c(N_c-1)}{2}\) cosine distance values for country \(c\).

We use cosine distance because it compares the shape of two personality profiles, that is, the relative pattern of high and low facets, and is invariant to proportional rescaling of a profile. This reduces, although it does not remove, sensitivity to differences in overall response level, such as acquiescent or extreme responding, which vary across cultures.

Given the highly positively skewed distribution of \(\mathbf{D}_c\), we focus on the positive distance values. Zero distances are excluded because their logarithm is undefined. We define the set of positive cosine distances as:
\begin{equation} \label{equ2}
\centering
\mathbf{D}^+_c = \{d \in \mathbf{D}_c \mid d>0\}
\end{equation}

To further address the skewness and normalize the distribution, we apply a natural logarithmic transformation to the positive cosine distances:
\begin{equation} \label{equ3}
\centering
\mathbf{D}^{\prime}_c = \{\ln(d) \mid d \in \mathbf{D}^+_c\}
\end{equation}

The logarithmic transformation compresses the range of distance values, reducing the impact of extreme values and facilitating a more balanced distribution analysis. Since the logarithm of distance values less than one yields negative results, we take the inverse of the absolute median of the transformed distribution \(\mathbf{D}^{\prime}_c\) as $\Psi$-GPDI, denoted \(D_c\) for each country \(c\).
\begin{equation} \label{equ4}
\centering
D_c = \frac{1}{|\operatorname{median}(\mathbf{D}^{\prime}_c)|}
\end{equation}

Because the index is calculated from cosine distance rather than similarity, larger typical distances (more dissimilar personality profiles) give higher values, so a higher $\Psi$-GPDI indicates greater personality diversity within a country.

\subsection*{Multifactor analysis of three factors}

The gross domestic product (GDP) data used in this research are sourced from the World Bank Group's World Development Indicators database, accessible at their website~\cite{worldbank_wdi}. For each country we use GDP per person employed in constant 2021 PPP international dollars (series SL.GDP.PCAP.EM.KD), averaged across the years 2005 to 2023. The arithmetic mean is taken over annual levels, and the resulting country average is then log-transformed for analysis. Ten countries lack estimates for one or more years in this window, so their averages are computed over 17 or 18 years rather than the full 19. Of the 135 countries in the $\Psi$-GPDI sample, six (Venezuela, Cuba, Yemen, Andorra, Antigua and Barbuda, and the British Virgin Islands) have no World Bank estimate of GDP per person employed and are excluded from the correlation analysis. A further 13 countries (Afghanistan, Nepal, Bosnia and Herzegovina, Georgia, North Macedonia, Cambodia, Mauritius, Barbados, Cabo Verde, Belize, Maldives, Fiji, and Rwanda) lack expropriation risk data and are excluded from the multifactor analysis, giving n = 116.

Acemoglu et al.~\cite{acemoglu2001colonial} use average protection against expropriation risk as a proxy for institutional quality. The measure is derived from assessments published by Political Risk Services and averaged over 1985 to 1995, and it captures the security of private property rights against government appropriation. Higher values indicate more secure property rights. For brevity we refer to this variable as expropriation risk throughout.

Alesina et al.~\cite{alesina2016birthplace} define immigrant diversity as the heterogeneity of immigrants based on their country of origin within a given population. The authors constructed a dataset on immigrant diversity using 2000 data from the OECD and the World Bank. In our multifactor analysis, we incorporate expropriation risk, immigrant diversity, and $\Psi$-GPDI as explanatory variables to evaluate their contributions to GDP. These factors are selected due to their relevance in capturing institutional quality, demographic diversity, and psychological diversity, each plausibly related to economic development.

The variables are measured over different periods: expropriation risk over 1985--1995, immigrant diversity around 2000, personality between 2001 and 2022, and GDP per person employed over 2005--2023, so the explanatory variables precede or overlap the outcome window. Institutional quality and the origin composition of immigrant populations change slowly, and we treat national personality diversity measured over 2001--2022 as a persistent characteristic of each population, although generational change in personality means that this assumption should be tested with time-varying data. The design therefore compares persistent national characteristics and does not establish temporal precedence or causal direction.

We employ multifactor regression analysis to examine the relationship between GDP \(Y\) and three independent variables: expropriation risk \(X_1\), immigrant diversity \(X_2\) and $\Psi$-GPDI \(X_3\). To ensure comparability of the coefficients and to facilitate interpretation, all independent variables are standardized (mean of zero, standard deviation of one) before being included in the regression models. We specify the multiple linear regression model as follows:
\begin{equation} \label{equ5}
\centering
Y = \beta_0 + \beta_1 X_1 + \beta_2 X_2 +\beta_3 X_3 + \epsilon
\end{equation}
where \(\beta_0\) is the intercept and \(\beta_1\), \(\beta_2\), and \(\beta_3\) are the regression coefficients. \(\epsilon\) is the error term.

To assess the severity of multicollinearity among the independent variables, we employ the Variance Inflation Factor (VIF). It is calculated using the formula:
\begin{equation} \label{equ6}
\centering
VIF_i = \frac{1}{1-R^2_i}
\end{equation}

where \(R^2_i\) is the coefficient of determination from regressing the \(i\)-th independent variable against all other independent variables. The calculated variance inflation factors of expropriation risk, immigrant diversity and $\Psi$-GPDI are 1.26, 1.14 and 1.12, respectively. Since all VIF values are below the commonly accepted threshold of 5, we conclude that multicollinearity does not pose a significant issue in our models.
We then test different regression models with various combinations of the independent variables to isolate their individual effects and assess their combined influence on GDP. By comparing the adjusted \(R^2\)  values across the different models, we assess the incremental explanatory power contributed by each independent variable.

\clearpage

\renewcommand{\figurename}{Fig.}
\renewcommand{\thefigure}{S\arabic{figure}}
\renewcommand{\tablename}{Table}
\renewcommand{\thetable}{S\arabic{table}}
\setcounter{figure}{0}
\setcounter{table}{0}

\section*{Supplementary Information}

\subsection*{Between-country structure of national personality profiles}

\begin{figure}[t!]
\centering
  {\includegraphics[width=\linewidth, keepaspectratio]{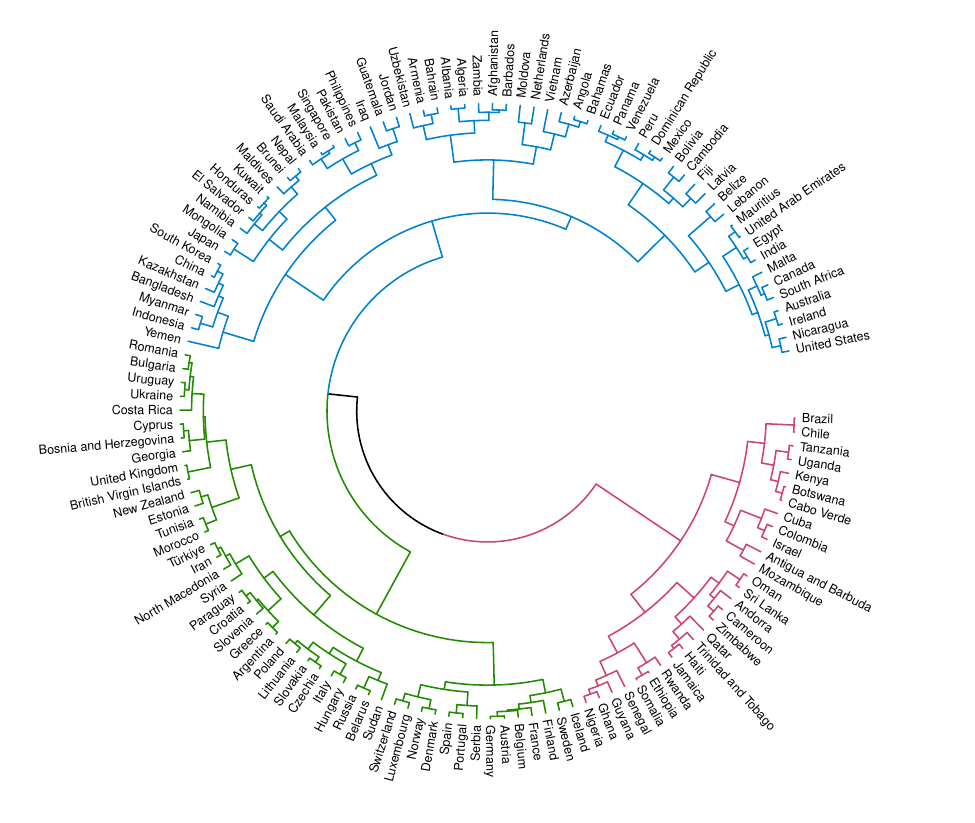}}
  \caption{\textbf{$|$ National personality profiles cluster partly, but not wholly, along regional lines.} Hierarchical clustering of 135 countries by their national personality profile, defined as the median score on each of the 30 Big Five facets (n = 760,242 individuals). Countries joined nearer the rim have more similar profiles, and branch colors mark the three top-level clusters (blue, green and red) described in SI Appendix. Some pairings follow geography and shared history (Spain and Portugal, Austria and Germany, South Korea and Japan), others link distant countries (New Zealand and Estonia, Greece and Argentina), and no cluster corresponds to a single world region. The figure compares typical national profiles. It does not measure the diversity of personalities within each country, which main text Figs.~\labelcref{fig:icc,fig:pairwise_distributions,fig:map} address.}\label{fig:dendrogram}
\end{figure}

For each of the 135 countries, we construct a 30-dimensional profile from the median scores of the Big Five personality facets and apply hierarchical clustering to identify countries with similar combinations of personality traits (SI Appendix, Fig.~\ref{fig:dendrogram}). This analysis captures \textit{between-country similarity} in multidimensional personality profiles rather than the degree of personality diversity within individual countries. Previous cross-national clustering based on broad Big Five domains identified meaningful similarities among 36 countries~\cite{allik2004toward}; we extend this approach to 30 facets and 135 countries.

The dendrogram reveals substantial cross-national structure. At the highest level, three broad clusters emerge. The blue cluster, the largest and most geographically heterogeneous, encompasses countries across Asia, the Middle East, Africa, and the Americas, including the United States, Canada, Australia, China, Japan, India, and Mexico. The green cluster is predominantly composed of European countries, including the United Kingdom, France, Germany, Spain, and Italy, together with several Central and Eastern European countries and a smaller number from other regions. The red cluster contains primarily countries from Africa, Latin America, the Caribbean, and the Middle East, including Kenya, Tanzania, Botswana, Brazil, Chile, Colombia, and Oman.

At finer levels of the hierarchy, several geographically or culturally proximate countries exhibit similar personality profiles, including Spain and Portugal, Austria and Germany, Brazil and Chile, and South Korea and Japan. Geographic proximity is not sufficient to explain the observed structure, however. Cross-regional pairings, such as New Zealand and Estonia, Paraguay and Croatia, and Greece and Argentina, indicate that countries separated geographically may nevertheless exhibit similar combinations of personality facets. These patterns should not be interpreted as evidence of common causal mechanisms; rather, they identify cross-national similarities that warrant further investigation in relation to cultural, historical, demographic, institutional, and socioeconomic characteristics.

\subsection*{U.S. personality archetypes (PLATYPUS)}

The U.S. sample comprised 446,783 individuals, each represented by scores across the 30 Big Five personality facets. To reduce the computational burden of hierarchical clustering, we randomly selected 10\% of the U.S. observations using a fixed random seed (\texttt{random\_state = 42}), resulting in a sample of 44,678 individuals. Prior to clustering, the 30-dimensional personality profiles were projected into a two-dimensional representation using t-distributed stochastic neighbor embedding (t-SNE), with a fixed random seed (\texttt{random\_state = 123}) to ensure reproducibility. Hierarchical agglomerative clustering was then applied to the resulting two-dimensional representations using Ward's linkage criterion and Euclidean distance. Ward's method iteratively merges clusters so as to minimize the increase in within-cluster variance.

\begin{figure*}[t!]
\centering
  {\includegraphics[width=\textwidth, keepaspectratio]{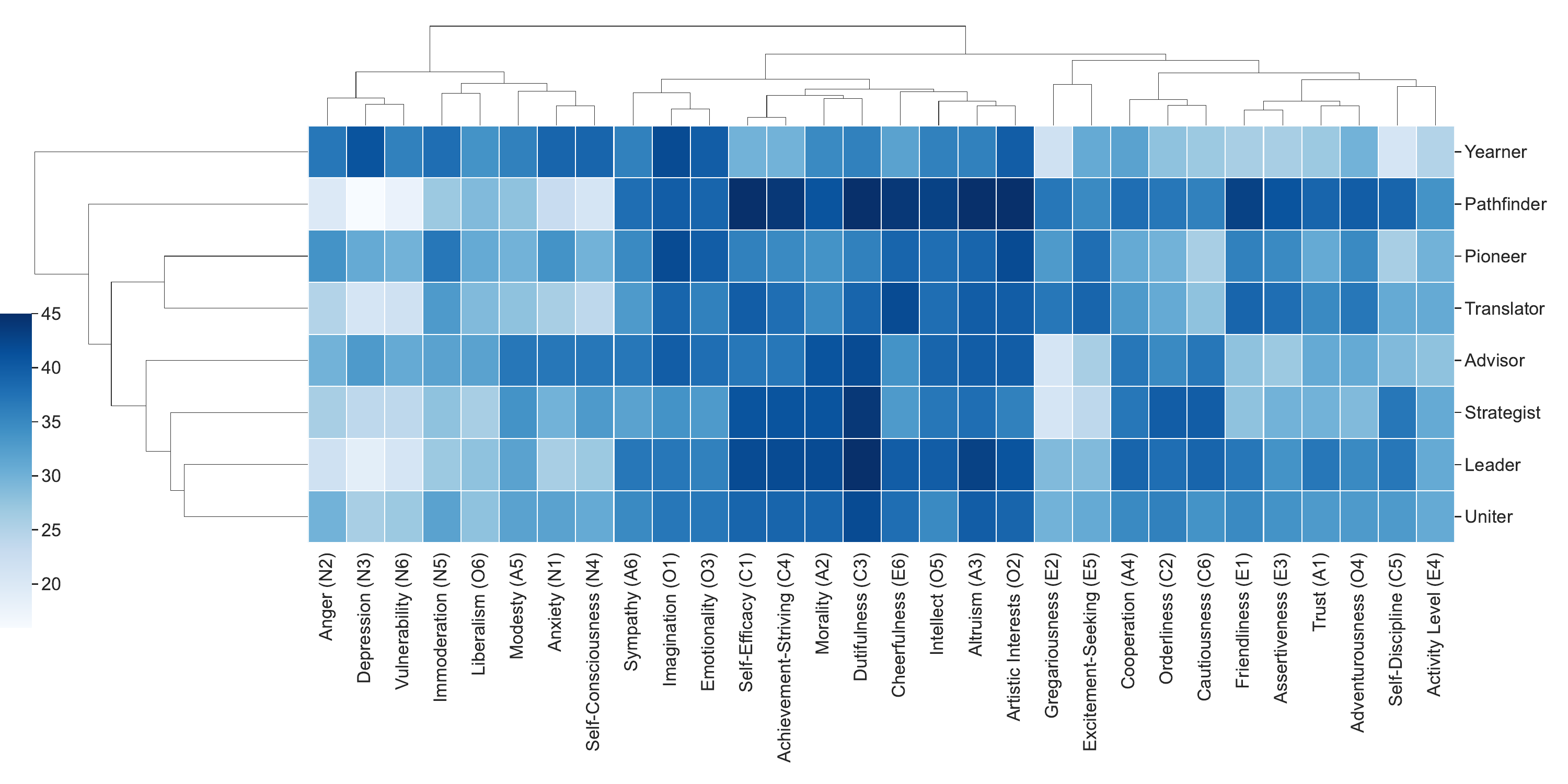}}
  \caption{\textbf{$|$ Within one country, people fall into archetypal personality types with different economic roles.} Using U.S. respondents as an illustrative case (n = 44,678, a random 10\% sample), hierarchical clustering identifies eight archetypes, each a distinct combination of high and low facets rather than a single dominant trait. The heatmap shows the median score on each of the 30 Big Five facets (columns) for each archetype (rows), with darker cells indicating higher scores (facet score range 10 to 50); the code after each facet name gives its Big Five domain and facet number. Dendrograms show how the archetypes group into broader families (left) and which facets vary together across archetypes (top). The eight archetypes, \textbf{P}athfinder, \textbf{L}eader, \textbf{A}dvisor, \textbf{T}ranslator, \textbf{Y}earner, \textbf{P}ioneer, \textbf{U}niter and \textbf{S}trategist, form the PLATYPUS typology (SI Appendix, Table~\ref{tab:platypus}). They differ in economically relevant ways as well as psychologically: predicted entrepreneurial propensity in the Pathfinder and Leader archetypes is about twice that of the other six (SI Appendix, Fig.~\ref{fig:prob}). The figure illustrates that a single national population contains a structured mix of personality types; other countries may show different archetypes, combinations or proportions.}\label{fig:us_heatmap}
\end{figure*}

To determine an appropriate number of clusters, we evaluated candidate solutions ranging from 3 to 39 clusters using two complementary internal validation measures, the Davies–Bouldin index and the silhouette coefficient. Lower Davies–Bouldin values indicate more compact and better-separated clusters, whereas higher silhouette coefficients indicate stronger separation between clusters. The inverse Davies–Bouldin scores and silhouette coefficients were standardized separately and averaged to construct a combined clustering-quality index. Inspection of these measures identified several candidate solutions, including 8, 13, and 22 clusters. We selected the eight-cluster solution because it provided the strongest balance of cluster compactness and separation while maintaining an interpretable representation of the underlying personality structure. To characterize each cluster, we calculated the median score for each of the 30 facets among individuals assigned to that cluster; these profiles are visualized in SI Appendix, Fig.~\ref{fig:us_heatmap} and summarized in SI Appendix, Table~\ref{tab:platypus}. Because t-SNE is designed primarily for visualization and can distort neighborhood structure, the archetypes should be regarded as descriptive. Their stability across random seeds and subsamples, and under clustering in the original standardized 30-dimensional space, remains to be tested.

\begin{figure}[!t]
\centering
  {\includegraphics[width=\linewidth, keepaspectratio]{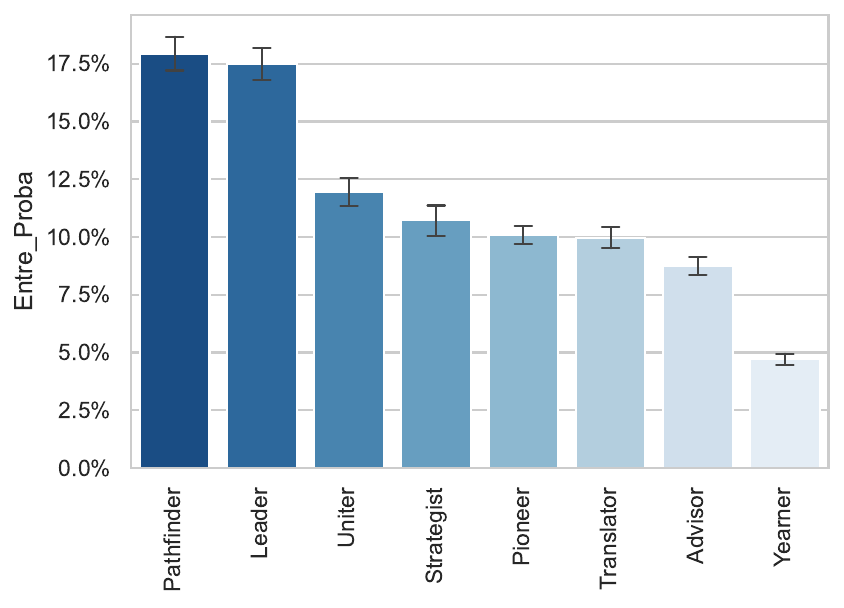}}
  \caption{\textbf{$|$ Two of the eight U.S. personality types have about twice the predicted entrepreneurial propensity of the others.} Bars show, for each of the eight U.S. personality clusters (SI Appendix, Fig.~\ref{fig:us_heatmap} and Table~\ref{tab:platypus}), the share of individuals predicted to have an entrepreneurial personality profile, sorted from highest to lowest; error bars show 95\% confidence intervals. Predictions come from a machine learning model based on the 30 Big Five facets (88\% accuracy), developed in~\cite{mccarthy2023impact}. Pathfinders and Leaders (about 17\% each) are roughly twice the average of the other six clusters (about 9\%), and Yearners are lowest (about 5\%). The types therefore differ in an economically relevant capacity, not only in psychological profile. A lower rate does not imply lower economic value, since the other types may suit different roles (SI Appendix, Table~\ref{tab:platypus}).}\label{fig:prob}
\end{figure}

To examine the functional implications of these clusters, we applied a Big Five 30-facet machine-learning predictor of entrepreneurial personality with 88\% accuracy, developed in a prior study~\cite{mccarthy2023impact}, to estimate the share of individuals in each cluster likely to possess personality traits associated with successful entrepreneurs. Two of the eight clusters, Pathfinders and Leaders, contain a markedly higher proportion of such individuals, at roughly 17\% each, close to double the rate in the remaining six clusters (SI Appendix, Fig.~\ref{fig:prob}).

\begin{table*}[t!]
\centering
\vspace{0.2cm} 
\centering
\makebox[\textwidth][c]{
  \begin{tabular}{llp{5cm}p{4cm}}
    \toprule
    Type & Archetype Description & Key Big Five Personality Facets & Economic Functions \\
    \midrule
    \textbf{P}athfinder & Proactive Collaborators & High in Agreeableness (Trust); Openness (Adventurousness) and Extraversion (Activity Level) & Build cohesive teams and foster an inclusive, collaborative culture. \\
    \textbf{L}eader & Charismatic Leaders & High in Agreeableness (Cooperation and Morality) and Conscientiousness (Dutifulness). & Foster collaboration and drive sustainable growth. \\
    \textbf{A}dvisor & Empathetic Guides & High in Conscientiousness (Dutifulness) and Agreeableness (Morality and Altruism)  & Maintain social infrastructure, essential for productivity. \\
    \textbf{T}ranslator & Optimistic Connectors & High in Extraversion (Cheerfulness, Excitement-Seeking, Gregariousness) & Create cohesive social and professional networks. \\
    \textbf{Y}earner & Emotional Idealists & High in Openness (Imagination, Emotionality) and Neuroticism (Anxiety) & Offer distinctive perspectives on human experiences. \\
    \textbf{P}ioneer & Creative Empaths & High in Openness (Artistic Interests) and Low Conscientiousness (Cautiousness) & Foster innovation and social well-being through creativity and spontaneity. \\
    \textbf{U}niter & Balanced Dependables & Moderate in all Big Five traits, showing versatility and flexibility & Bridge gaps between creativity and practical implementation. \\
    \textbf{S}trategist & Practical Realists & High in Conscientiousness (Orderliness, Cautiousness), Low in Agreeableness (Sympathy) & Thrive in roles requiring precision and careful planning. \\
    \bottomrule
  \end{tabular}
}
\caption{Eight distinct personality types were identified within the United States (n = 44,678) using Big Five personality traits. Each type is described by an archetype, key personality facets, and its economic function, illustrating the diverse roles individuals play in fostering innovation, collaboration, and productivity.}\label{tab:platypus}
\end{table*}

\begin{figure}[!t]
\centering
  {\includegraphics[width=\linewidth, keepaspectratio]{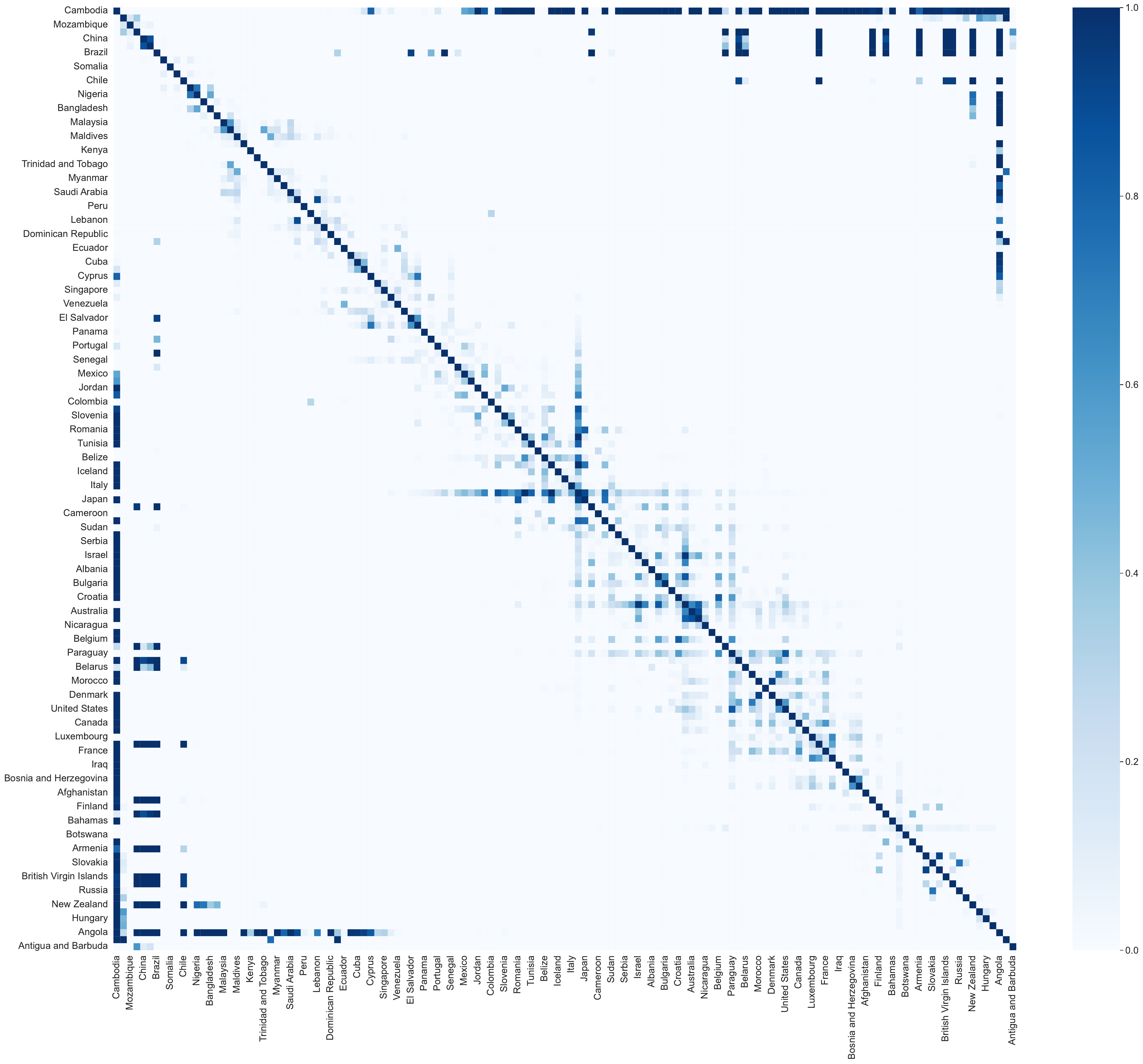}}
  \caption{\textbf{$|$ Most countries' distributions of personality distances differ significantly from one another.} Each cell shows the $p$-value of a two-sample Kolmogorov--Smirnov test comparing two countries' distributions of log pairwise cosine distances (main text Fig.~\ref{fig:pairwise_distributions}). Countries are ordered by $\Psi$-GPDI from lowest (top, left) to highest (bottom, right), and every second country is labeled. Light cells ($p$ near 0) mark pairs whose distributions differ significantly; dark cells mark pairs that cannot be distinguished, and the diagonal compares each country with itself. 93.53\% of the 9,045 country pairs differ significantly. Because pairwise distances within a country share individuals, these p-values are descriptive. Apart from a few countries, such as Cambodia and Angola, indistinguishable pairs lie close to the diagonal among countries with similar $\Psi$-GPDI, consistent with the index capturing the main differences between national distributions.}\label{fig:ks_heatmap}
\end{figure}

\begin{figure}[t!]
\centering
  {\includegraphics[width=\linewidth, keepaspectratio]{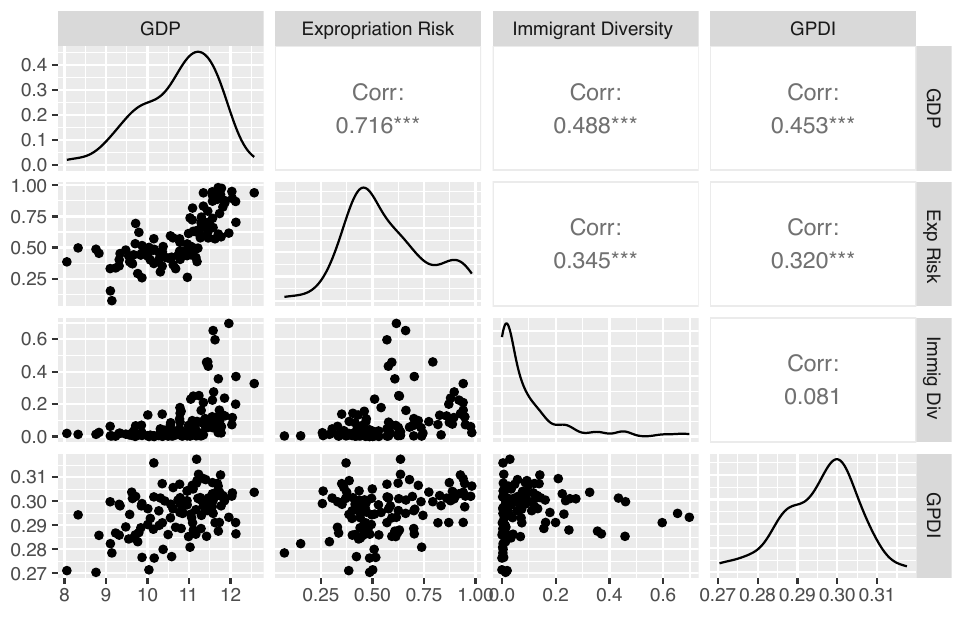}}
  \caption{\textbf{$|$ Personality diversity tracks national income but is largely independent of immigrant diversity.} Scatterplots and correlation matrix illustrating the relationships between GDP per person employed, $\Psi$-GPDI, immigrant diversity, and expropriation risk across 116 countries with complete data on all four measures. $\Psi$-GPDI is moderately correlated with GDP ($r = 0.453$), comparable to immigrant diversity ($r = 0.488$) and weaker than expropriation risk ($r = 0.716$). $\Psi$-GPDI is essentially uncorrelated with immigrant diversity ($r = 0.081$, not significant), indicating that it captures variation in national populations that is distinct from migration. Asterisks denote statistical significance: *** $p<0.001$.}\label{fig:corr}
\end{figure}

\begin{figure}[!t]
\centering
  {\includegraphics[width=\linewidth, keepaspectratio]{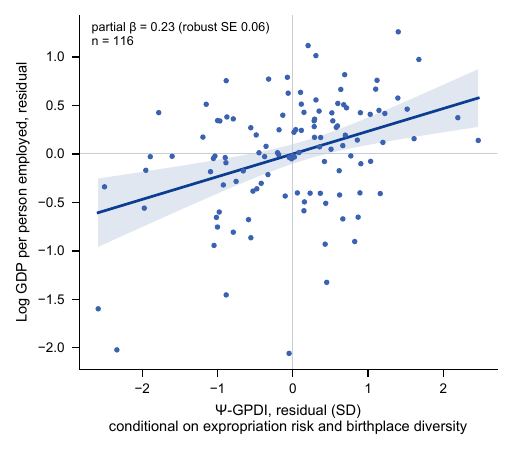}}
  \caption{\textbf{$|$ Personality diversity remains associated with national income after conditioning on institutions and birthplace diversity.} Partial regression (added-variable) plot. The horizontal axis shows $\Psi$-GPDI (in standard deviations) and the vertical axis log GDP per person employed, each after removing the variation explained by expropriation risk and birthplace diversity. Each point is a country ($n = 116$). The line is the fitted partial relationship (partial $\beta = 0.23$, robust SE $= 0.06$) and the shaded band indicates the uncertainty around it.}\label{fig:gpdi_residual}
\end{figure}

\begin{figure}[!t]
\centering
  {\includegraphics[width=\linewidth, keepaspectratio]{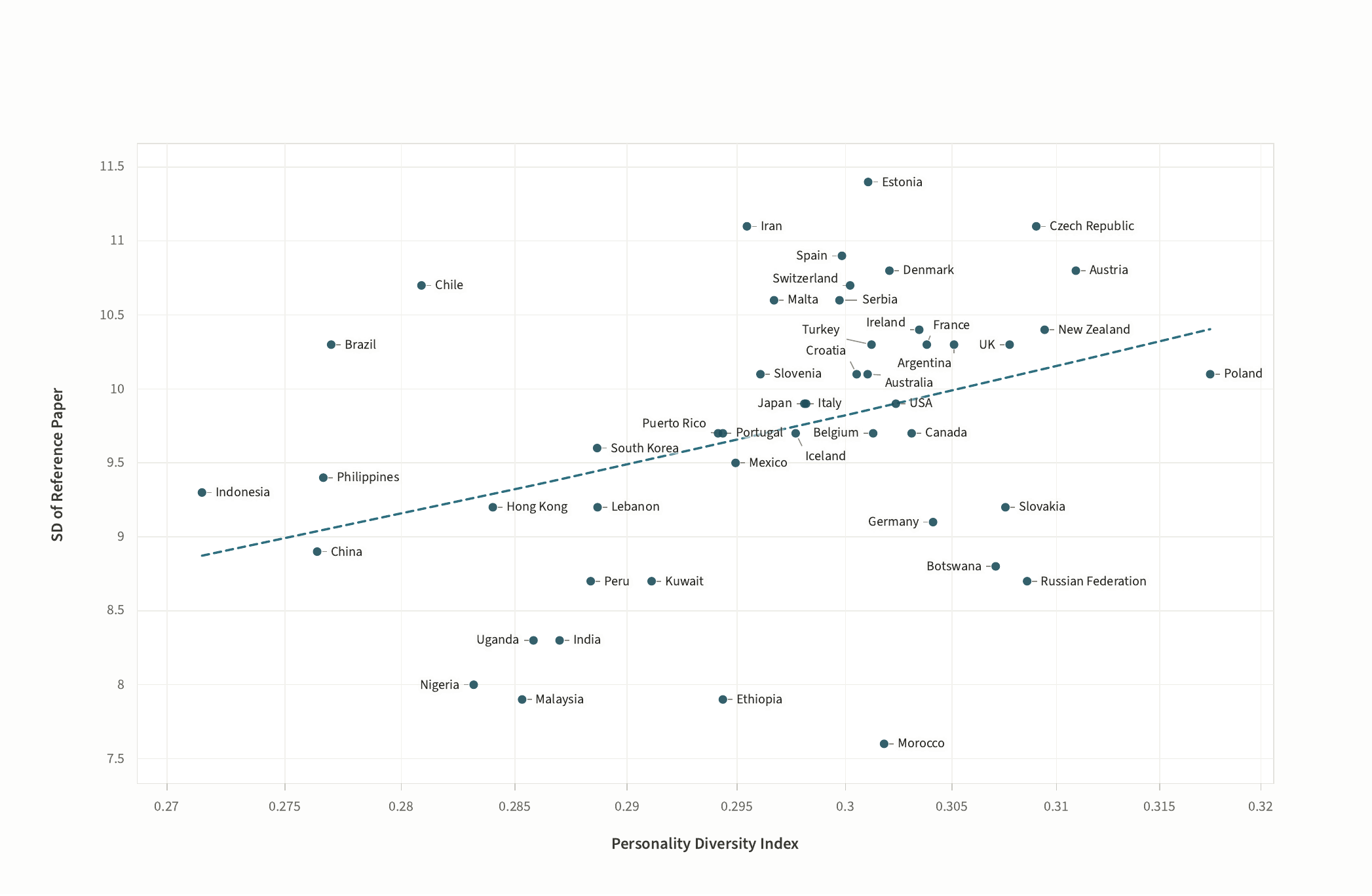}}
  \caption{\textbf{$|$ $\Psi$-GPDI is modestly associated with an independent, observer-rated measure of within-country personality variation.} Relationship between country-level $\Psi$-GPDI scores derived from self-reported IPIP-NEO personality data and the standard deviations of personality scores reported by McCrae et al.~\cite{mccrae2005personality}, based on independent observer ratings (averaged across the 30 facets within five domains). Each point represents a country included in both datasets (n = 48), and the dashed line represents the fitted linear regression. The positive and statistically significant association ($R^2 = 13.26\%$, $p = 0.0109$) indicates convergence between the two independently derived measures of within-country personality variation, providing modest convergent evidence for $\Psi$-GPDI. The modest $R^2$ is expected because the two measures differ in instrument, rater and statistic; they are related but not interchangeable.}\label{fig:index_validation}
\end{figure}

\begin{figure}[!t]
\centering
  {\includegraphics[width=\linewidth, keepaspectratio]{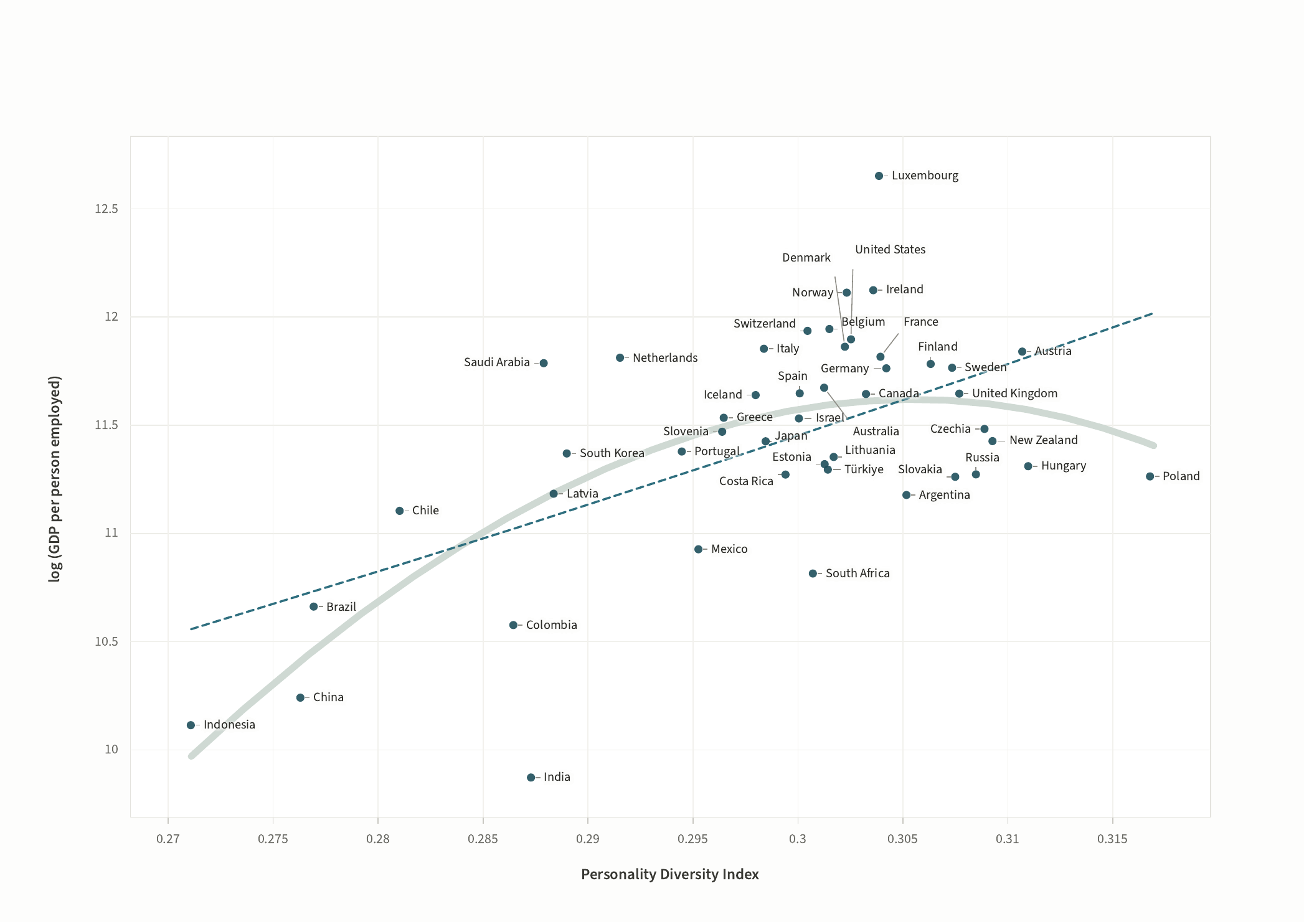}}
  \caption{\textbf{$|$ The association between personality diversity and income holds among OECD and G20 countries.} GDP per person employed and $\Psi$-GPDI for the 46 countries in the union of the OECD and G20 memberships that are present in the sample. GDP per person employed is averaged over 2005--2023 and expressed in constant 2021 PPP international dollars, and values on the vertical axis are natural logarithms. The blue dashed line is a linear fit ($R^2 = 32.47\%$) and the grey curve a quadratic fit ($R^2 = 41.53\%$). The association is stronger within this group of mostly higher-income economies than in the full sample (main text Fig.~\ref{fig:regression}), indicating that it does not arise only from contrasts between rich and poor countries. As in the full sample, the quadratic fit flattens at the highest values of $\Psi$-GPDI.}\label{fig:combined_regression_a}
\end{figure}

\begin{figure}[!t]
\centering
  {\includegraphics[width=\linewidth, keepaspectratio]{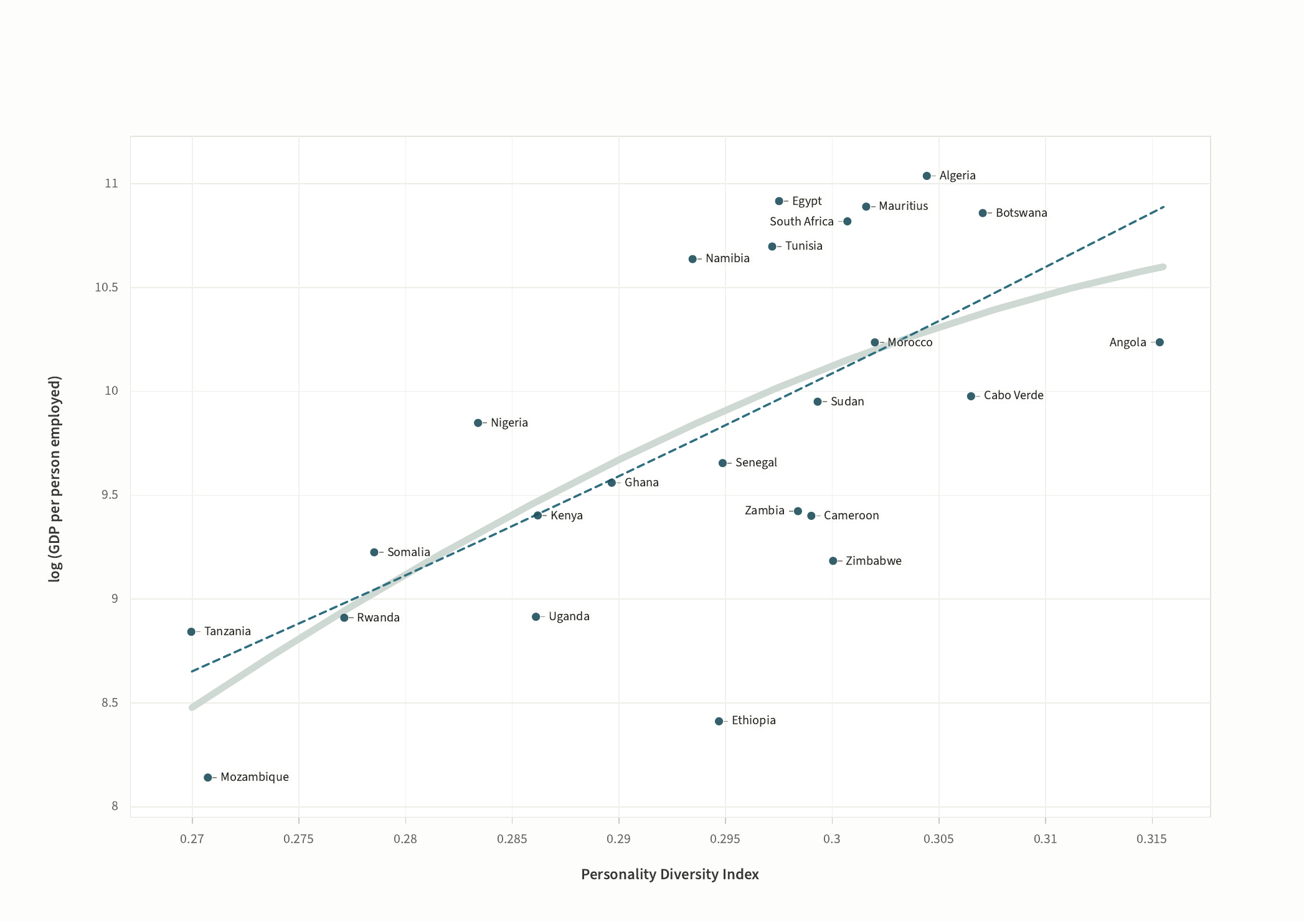}}
  \caption{\textbf{$|$ The association between personality diversity and income also holds among African countries.} GDP per person employed and $\Psi$-GPDI for the 24 African countries in the sample. GDP per person employed is averaged over 2005--2023 and expressed in constant 2021 PPP international dollars, and values on the vertical axis are natural logarithms. The blue dashed line is a linear fit ($R^2 = $ 44.85\%) and the grey curve a quadratic fit ($R^2 = $ 46.05\%). These countries occupy a lower income range than the OECD and G20 group (Fig.~\ref{fig:combined_regression_a}), so the positive slope here shows that the relationship holds across very different levels of development, not only among high-income economies.}\label{fig:combined_regression_b}
\end{figure}

\begin{figure}[!t]
\centering
  {\includegraphics[width=\linewidth, keepaspectratio]{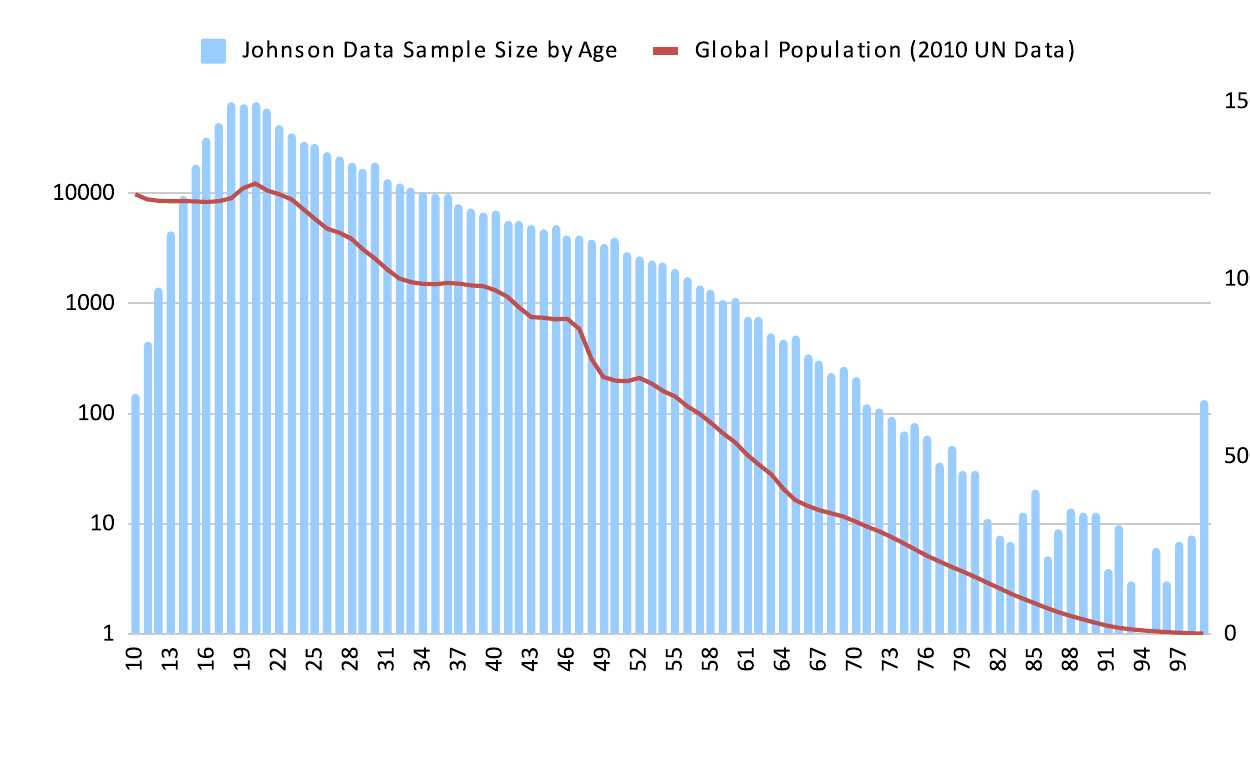}}
  \caption{\textbf{$|$ The age profile of the sample broadly follows the global age distribution, though the sample is younger.} Bars show the number of respondents at each year of age (left axis, logarithmic scale); the red line shows global population by age in 2010 from the United Nations (right axis, tens of millions). The two are positively correlated across ages ($r = 0.64$), indicating that the sample spans the age range in broadly similar relative proportions. Respondents in their late teens and twenties are over-represented (sample median age 22, compared with a global median of about 28 in 2010), which should be kept in mind when generalizing to national populations.}\label{fig:age_population}
\end{figure}

\begin{figure}[!t]
\centering
  {\includegraphics[width=\linewidth, keepaspectratio]{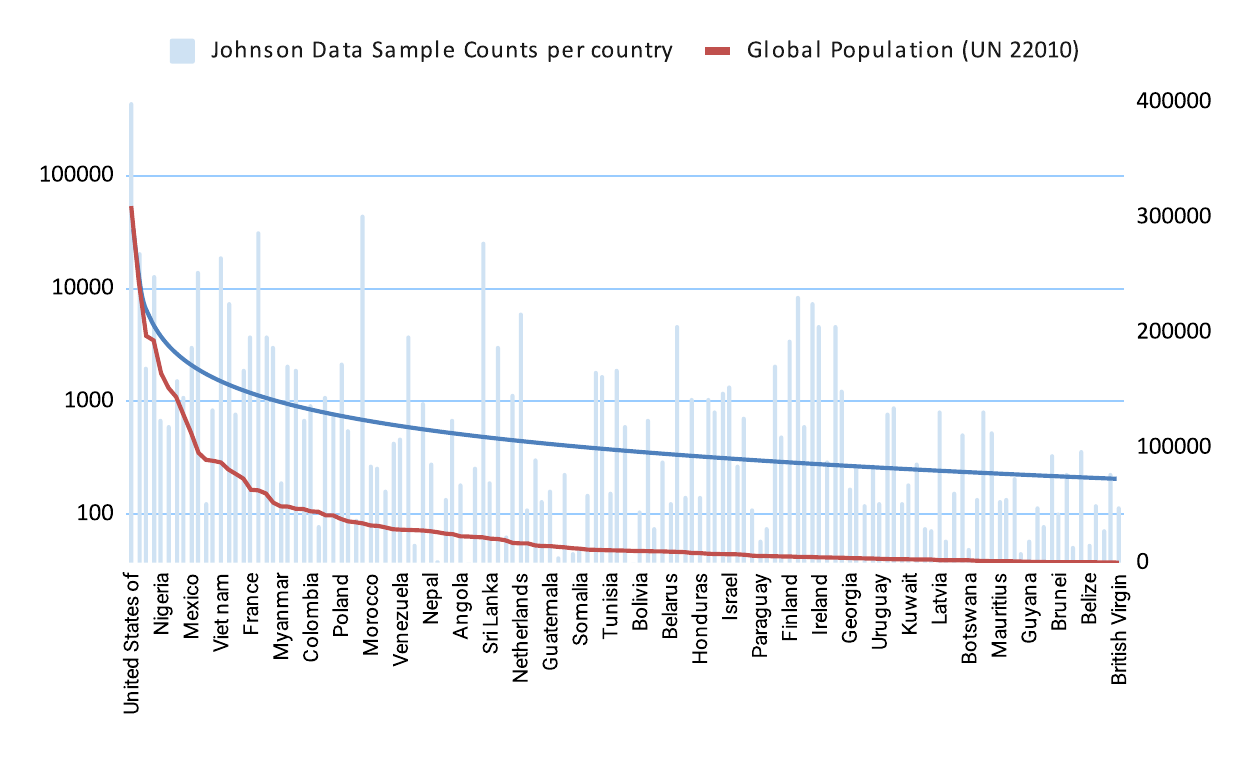}}
  \caption{\textbf{$|$ Sample sizes by country broadly track national populations.} Bars show the number of respondents from each country (left axis, logarithmic scale), with countries ordered by population; the red line shows national population in 2010 from the United Nations (right axis, thousands), and the blue curve is a logarithmic trend fitted to the sample counts. China and India are omitted for clarity. Sample size and population are positively correlated ($r = 0.54$), so more populous countries generally contribute more respondents. Bars well above or below the trend identify countries that are over- or under-represented relative to their population.}\label{fig:country_population}
\end{figure}

\clearpage

\printbibliography

\end{document}